%% file: acl_latex.tex
\documentclass[11pt]{article}

\usepackage[final]{acl}

\usepackage{times}
\usepackage{latexsym}

\usepackage[T1]{fontenc}

\usepackage[utf8]{inputenc}

\usepackage{microtype}

\usepackage{inconsolata}

\usepackage{amsmath} 
\usepackage{graphicx}
\usepackage{xcolor}
\usepackage{hyperref}
\usepackage{url}
\usepackage{pifont}
\usepackage{xspace}
\usepackage{booktabs}
\usepackage{caption}
\usepackage{makecell}
\usepackage{wrapfig}
\usepackage{multirow}
\usepackage{colortbl}
\usepackage{textcomp}
\usepackage{enumitem}

\newcommand{\cmark}{\ding{51}}  
\newcommand{\xmark}{\ding{55}}  

\newcommand{\bench}{\textsc{COIN-Bench}\xspace}

\newcommand{\tree}{\textsc{COIN-Tree}\xspace}
\newcommand{\rag}{\textsc{COIN-RAG}\xspace}

\definecolor{lightgray}{gray}{0.95}
\title{COINBench: Moving Beyond Individual Perspectives to Collective Intent Understanding}

\author{Xiaozhe Li$^{1}$\footnotemark[1] \quad  Tianyi Lyu$^{1}$\footnotemark[1] \quad Siyi Yang$^{1}$ \quad Yizhao Yang$^1$ \quad Yuxi Gong$^1$\\ \bfseries  Jinxuan Huang$^{1}$ \quad Ligao Zhang$^3$ \quad Zhuoyi Huang$^{2,3}$ \quad Qingwen Liu$^1$\footnotemark[2] \\ Tongji University$^1$ \quad Stanford University$^2$ \quad CurrentsAI Research$^3$}

\begin{document}
\twocolumn[{%
\renewcommand\twocolumn[1][]{#1}%
\maketitle
\vspace{-10mm}

\begin{center}
    \centering
    \captionsetup{type=figure}
    \includegraphics[width=1\linewidth]{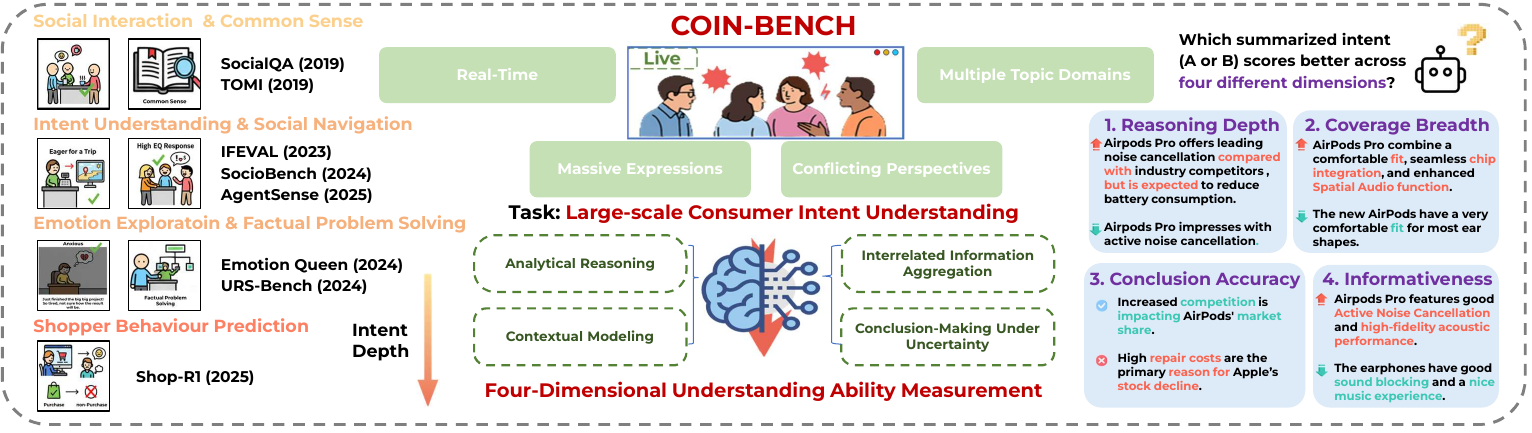}
    \caption{\textbf{Motivation for \bench.} Existing benchmarks predominantly rely on individual perspectives and synthetic data, which fail to assess model performance in deciphering complex, real-world collective intent. We propose \bench, a large-scale, live-updating benchmark that evaluates collective intent understanding across four key dimensions: depth, breadth, accuracy, and informativeness.}
    \label{fig:spotligh}
    \vspace{3mm}
\end{center}%
}
]
\footnotetext[1]{* Equal Contribution.}
\footnotetext[2]{$^{\dag}$ Corresponding Author.}
\input{section/abs}
\input{section/0intro}
\input{section/1method}
\input{section/3experiment}

\input{section/2related_work}
\input{section/4conclusion}
\section*{Limitations}
Due to resource constraints, we were unable to include additional state-of-the-art models such as the Gemini 3 series, OpenAI GPT-5.2, and Claude 4.5 in our experiments. 
\bibliography{custom}

\appendix

\input{section/appendix}

\label{sec:appendix}

\end{document}

%% file: section/abs.tex
\begin{abstract}
Understanding human intent is a high-level cognitive challenge for Large Language Models (LLMs), requiring sophisticated reasoning over noisy, conflicting, and non-linear discourse. While LLMs excel at following individual instructions, their ability to distill \textbf{Collective Intent}—the process of extracting consensus, resolving contradictions, and inferring latent trends from multi-source public discussions—remains largely unexplored. To bridge this gap, we introduce \textsc{\bench}, a dynamic, real-world, live-updating benchmark specifically designed to evaluate LLMs on collective intent understanding within the consumer domain. Unlike traditional benchmarks that focus on transactional outcomes, \textsc{\bench} operationalizes intent as a \textbf{hierarchical cognitive structure} (from explicit scenarios to deep causal reasoning). We implement a robust evaluation pipeline that synergizes rule-based method with an LLM-as-the-Judge approach. This framework incorporates \textsc{\tree} for hierarchical cognitive structuring and retrieval-augmented verification (\textsc{\rag}) to ensure expert-level precision in analyzing raw, collective human discussions. An extensive evaluation of 20 state-of-the-art LLMs across four dimensions—\textit{depth, breadth, informativeness, and correctness}—reveals that while current models handle surface-level aggregation, they struggle with the analytical depth required for complex intent synthesis. \textsc{\bench} establishes a new standard for advancing LLMs from passive instruction-followers to expert-level analytical agents capable of deciphering the collective voice of the real world. See our project page on \href{https://github.com/Virtu-llc/COINBench.github.io}{\bench}.
\end{abstract}

%% file: section/0intro.tex
\section{Introduction}
\label{section:introduction}
The emergence of Large Language Models (LLMs)~\cite{o1-system-card, grattafiori2024llama, guo2025deepseek} has catalyzed a fundamental paradigm shift in artificial intelligence, moving beyond mere text generation toward sophisticated reasoning and real-world problem-solving~\cite{tongyidr}. Current state-of-the-art models exhibit remarkable proficiency in structured, task-oriented environments, as evidenced by their performance on benchmarks such as SWE-bench~\cite{swebench} for software engineering, SPIDER2.0~\cite{lei2024spider} for SQL generation, and GAIA~\cite{mialon2023gaia} for tool-augmented reasoning. However, these evaluations predominantly focus on the model as a \textit{reasoning agent} interacting with a \textit{environment}, leaving a pivotal dimension of intelligence largely unexplored: understanding collective human intent. This task requires the sophisticated ability to distill coherent consensus from the noisy, fragmented, and often conflicting perspectives inherent in a population.

Real-world public discourse, such as community debates or consumer feedback, is characterized by "swarm-like" intelligence. Unlike linear instructions, these discussions involve an intricate interplay of divergent goals, emotional nuances, and implicit assumptions. To truly comprehend this environment, an LLM must go beyond parsing individual sentences; it must synthesize multi-source signals, resolve contradictions, and infer latent trends within an evolving discourse. This capability—acting as an expert-level analyst capable of navigating complex public opinion—represents a high-level cognitive frontier that current evaluation frameworks fail to capture.

To rigorously investigate this capability, we identify the \textbf{consumer domain} as an ideal microcosm for \textbf{Collective Intent}. Consumer discussions are far more than transactional records; they embody grounded yet complex human reasoning, rich in the noise and redundancy inherent to collective intelligence. While prior benchmarks have explored social or emotional dimensions~\cite{sap2019socialiqa, chen2024socialbench, mou2025agentsense}, they often rely on hand-crafted or semi-synthetic datasets that lack the "raw noise" of real-world discourse. Furthermore, recent consumer-focused benchmarks like Shop-R1~\cite{zhang2025shopr1} primarily treat intent as a surface-level prediction of shopper actions (e.g., purchase vs. non-purchase). We argue that such transactional outcomes overlook the deeper \textbf{Collective Intent}—the multi-layered cognitive structure distilled from a diverse population. There is a pressing need for a benchmark that shifts the focus from simple action prediction to the hierarchical synthesis of collective human voices.

To bridge this gap, we introduce \textsc{\bench} (Collective Intent Benchmark), a large-scale, dynamic, and live-updating framework designed to evaluate LLMs on their synthesis of real-world collective consumer intent. \textsc{\bench} encompasses over \textbf{200k opinions} across 9 major domains and 54 sub-categories, spanning 1,400+ products. Crucially, we depart from passive evaluation (e.g., multiple-choice questions) in favor of an \textbf{Active Probing Paradigm}. We task LLMs with acting as "meta-analysts" to reconstruct chaotic public discourse into structured questionnaires. This paradigm is underpinned by three core cognitive pillars:
\begin{itemize}[leftmargin=*, itemsep=2pt]
    \item \textbf{Consensus Extraction}: Designing valid questionnaires requires models to implicitly perform denoising and clustering, identifying salient topics amidst the noise.
    \item \textbf{\textsc{\tree} for Hierarchical Resolution}: Collective intent is inherently stratified. We map model outputs onto a hierarchical cognitive topology to strictly distinguish between surface-level aggregation (L1--L3) and deep causal reasoning or trend prediction (L4--L5).
    \item \textbf{\textsc{\rag} for Grounded Verification}: To ensure correctness, we employ a retrieval-augmented verification pipeline that penalizes hallucinations by ensuring all inferred consensus is strictly grounded in the raw discussion context.
\end{itemize}

Evaluating 20 diverse LLMs reveals a critical performance dichotomy: while reasoning-enhanced models (e.g., OpenAI o1/o3) demonstrate superior depth in penetrating complex cognitive layers, a significant gap remains in their ability to achieve broad and expert-level analytical synthesis. Our findings suggest that even the most advanced models struggle with the multifaceted nature of collective intent, highlighting substantial room for future development.

In summary, our contributions are as follows:
\begin{itemize}[leftmargin=*, itemsep=2pt]
    \item \textbf{A Large-Scale, Live Benchmark}: We introduce \textsc{\bench}, featuring 200k+ real-world consumer discussions across 1,400+ products. Its dynamic nature ensures real-time updates and mitigates data contamination.
    \item \textbf{Robust Active Evaluation Pipeline}: We propose the Active Probing Paradigm, which transforms traditional passive structural questioning into a process of model-led self-exploration and autonomous Q\&A. By integrating the \textsc{CoIn-Tree} with the \textsc{CoIn-RAG} verification framework, this pipeline ensures high-precision assessment, fundamentally shifting the evaluation focus from passive response generation to the active, structured reconstruction of collective intent.
    \item \textbf{Extensive Empirical Analysis}: We provide a comprehensive evaluation of 20 leading LLMs, revealing fundamental limitations in current models' ability to synthesize deep, multi-source human intent.
\end{itemize}

%% file: section/1method.tex
\section{BENCHMARK CONSTRUCTION}

\subsection{Data Curation}
Constructing a benchmark for collective intent requires a corpus that reflects the noisy, fragmented, and contradictory nature of real-world public discourse. We established an automated curation pipeline designed to retrieve high-density discussions while filtering out irrelevant noise. The process consists of three integrated stages:

\noindent\textbf{Hybrid Acquisition (Search \& Retrieval).} To maximize coverage, we employ a hybrid retrieval strategy combining semantic vector search with LLM-augmented API queries. 
For a given focal product (seed entity), the system first generates context-aware keywords using an LLM to expand the search space. 
It then retrieves discussion threads from diverse open-source web platforms. 
To ensure semantic relevance, we utilize vector embeddings to measure similarity between the seed topic and retrieved content, retaining only threads surpassing a strict similarity threshold.

\noindent\textbf{Denoising and Filtering.} 
Raw web data is inherently unstructured. We apply a cascade of quality filters: 
(1) \textit{Rule-based Filtering} removes low-information content (e.g., extremely short texts $<20$ chars); 
(2) \textit{Semantic Filtering} employs an LLM-based verifier to exclude discussions that drift from the core topic; 
(3) \textit{Temporal Alignment} prioritizes recent discussions to ensure the intent reflects current market realities. 
This pipeline ensures the final dataset preserves the complexity of swarm intelligence while maintaining high contextual fidelity.

\subsection{Data Statistics}
\input{table/data_statistic}

As detailed in Table~\ref{tab:bench} and Figure~\ref{fig:data}, \bench establishes a robust foundation for evaluating collective intent. The benchmark spans 9 primary domains and 54 sub-categories, covering over 1,400 focal products. 
Crucially, unlike datasets that focus on sparse queries, we enforce high information density: each product entry aggregates an average of 200 user comments, totaling over 200k individual opinions. This scale is essential for testing an LLM's ability to synthesize a "many-to-one" consensus from a large volume of diverse signals, detailed information in Appendix~\ref{apd:data_stat}.

\subsection{Evaluation Dimensions}  
To rigorously quantify the LLM's cognitive capability in collective intent synthesis, we categorize evaluation into four orthogonal dimensions:

\noindent\textbf{Depth (Cognitive Penetration).} 
This dimension measures the level of reasoning required to derive the intent. We introduce a five-level taxonomy (\textsc{CoIn-Tree} levels):
\begin{itemize}[leftmargin=*, nosep]
    \item \textbf{Observational Understanding (L1--L3):} The model identifies explicit information directly present in the text, such as \textit{Usage Scenarios} (L1), \textit{Product Aspects} (L2), and \textit{User Feelings} (L3).
    \item \textbf{Inferential Reasoning (L4--L5):} The model must synthesize dispersed information to perform advanced reasoning. This includes \textit{Comparative Analysis} against competitors (L4) and \textit{Predictive Tendency} regarding future improvements or market shifts (L5).
\end{itemize}
A high depth score indicates the model moves beyond surface summarization to expert-level insight.

\noindent\textbf{Breadth (Semantic Coverage).} 
Collective intent is multifaceted. This dimension evaluates the model's completeness in capturing the full spectrum of the crowd's voice. A high breadth score signifies that the model has successfully identified the "long tail" of user concerns—covering diverse usage scenarios and edge cases—rather than fixating on a single dominant topic.

\noindent\textbf{Correctness (Grounded Fidelity).} 
In swarm analysis, hallucination is the primary failure mode. This dimension evaluates whether the inferred intent is strictly grounded in the provided discussions. Using our \rag pipeline, we measure the factual alignment between the model's generated consensus and the raw crowd evidence, ensuring the output is free from bias and unverified extrapolation.

\noindent\textbf{Informativeness (Expressive Diversity).} 
An expert analyst should articulate insights with nuance. We assess this through \textit{Lexical Richness} (vocabulary range) and \textit{Semantic Distinctiveness} (lack of redundancy). High performance here indicates the model avoids generic boilerplate responses ("The product is good") and instead employs precise, varied language to capture the specific texture of the consumer intent.
\section{Methodology}

Evaluating \textbf{Collective Intent} presents a unique challenge: unlike factual QA, there is no single 'gold standard' sentence that captures the entirety of a chaotic public discussion. Passive evaluation methods (e.g., multiple-choice selection) fail to test whether a model can actively synthesize fragmented views into a coherent whole.

To address this, we operationalize evaluation through an \textbf{Active Probing Paradigm}. We task the LLM with acting as a 'Meta-Analyst' requiring it to \textbf{reconstruct} the unstructured crowd feedback into a structured \textbf{Questionnaire}. This design relies on the cognitive premise that to formulate a valid question-answer pair from noise, the model must successfully execute two high-level reasoning steps:
\begin{itemize}
    \item \textbf{Salience Identification (The Question):} By formulating a relevant question stem, the model demonstrates it can identify meaningful topics amidst noise.
    \item \textbf{Consensus Aggregation (The Answer):} By designing options that reflect the majority perspective, 
    the model demonstrates a precise grasp of intents within diverse opinion clusters.
\end{itemize}

To rigorously assess the quality of this reconstructed intent, we propose a robust evaluation framework combining topological analysis (\tree) and evidence-based verification (\rag), as illustrated in Figure~\ref{fig:pipeline}.

\input{Fig/pipeline}

\subsection{\tree: Hierarchical Cognitive Stratification}
\label{tree}


Moving beyond shallow keyword matching, \tree provides a \textbf{hierarchical intent graph} derived from real-world discussions to measure deep understanding. It allows to quantify the \textit{Hierarchic} of the model's understanding---distinguishing between surface-level observation and deep reasoning.

\noindent\textbf{Construction.} 
We operationalize collective intent as a five-level weighted hierarchical tree (\textsc{CoIn-Tree}), where the root denotes the focal product. The construction process involves two key stages: 
(i) \textbf{Branch Extraction:} We leverage an LLM to distill raw, unstructured discourse into standardized semantic branches based on a cognitive template (ranging from \textit{Usage Scenarios} to \textit{Future Tendencies}); 
(ii) \textbf{Node Aggregation:} Semantically redundant nodes across disparate branches are consolidated into high-weight nodes. 
The final weights are computed by integrating the frequency of node occurrences with real-world engagement metrics, such as view counts and upvotes, to reflect the popularity and consensus of specific viewpoints. Further technical details and the formal extraction prompt are provided in Figure~\ref{fig:pipeline} and Appendix~\ref{subsection: tree}.

\noindent\textbf{Surface Understanding (L1--L3):} Nodes at these levels capture explicit content directly observable in user discussions, including \textit{Usage Scenarios} (L1), specific \textit{Aspects} (L2), and \textit{User Feelings} (L3) of the product.

\noindent\textbf{Deep Reasoning (L4--L5):} These levels require the model to synthesize dispersed information. L4 (Comparison) represents comparative reasoning against competing products, while L5 (Tendency) represents predictive insights into future product improvements or market shifts.

\noindent\textbf{Evaluation (Lighten Tree Nodes).} 
We evaluate the Depth and Breadth of the LLM by projecting its generated questionnaire onto the \tree nodes. Specifically, the content of each generated question is extracted into a semantic branch and matched shallow-deep against the \tree nodes using a Sentence Transformer. Successful matches 'lighten' the corresponding nodes. The lightened sub-tree illustrates how well the model understands this human intent.

\noindent\textbf{Depth Score:} Calculated as the ratio of the total weight of lightened nodes at a specific level (L1--L5) to the total weight of all nodes at that level. The overall depth is the average  all five levels. A high score at L4/L5 indicates the model penetrated the ``deep structure'' of collective intent.

\noindent\textbf{Breadth Score:} Calculated as the sum of weights of all lightened nodes. This reflects the model's coverage of the multifaceted nature of the discussion.

\vspace{-1em}
\subsection{\rag: Evaluation of Grounded Correctness}
\label{rag}

In the context of swarm intelligence, 'correctness' implies fidelity to the crowd's consensus rather than external facts. To prevent the evaluation from favoring eloquent but hallucinated insights, we implement \rag, a retrieval-augmented verification pipeline acting as a ``Fact-Checker.''

\noindent\textbf{Construction.} 
\rag serves to ground the LLM's inferred intent in the raw evidence. We employ a two-stage retrieval process:

\noindent\textbf{Dual-Embedding:} User discussions are vectorized using both TF-IDF (for precise keyword matching) and \texttt{all-MiniLM-L6-v2} (for semantic nuance). This ensures retrieval captures both specific terminology and implicit sentiments.

\noindent\textbf{Joint Retrieval:} Key opinions extracted from the LLM's generated questionnaire are queried against the vector database to retrieve the top-$k$ most relevant raw discussion entries.

\noindent\textbf{Evaluation.} 
We employ a reasoning-based verification step. The retrieved raw discussions (Ground Truth Evidence) are compared against the LLM's generated intents (which are the question stems and given answers in the questionnaires). If the LLM's synthesized consensus aligns with the majority perspective found in the retrieved evidence, it is deemed correct. This pipeline strictly penalizes hallucinations that drift away from the authentic collective voice.

\vspace{-1em}
\subsection{Informativeness: Assessing Analytical Density and Diversity}
\label{sec:informativeness}

Beyond structural accuracy, expert-level intent synthesis must be both information-dense and analytically diverse. We evaluate the \textbf{Informativeness} of generated questionnaires to ensure that models capture the full nuance of collective discourse rather than producing generic or repetitive text.

\paragraph{Lexical Richness}
To reflect diverse population expression, models must employ precise and domain-relevant vocabulary. We measure lexical variety using the \textbf{Type–Token Ratio (TTR)}~\cite{johnson1944type} and phrase-level diversity using \textbf{Distinct-$n$}~\cite{li2016diversity}. Higher scores indicate effective abstraction of raw crowd language into richer analytical expression.

\paragraph{Semantic Redundancy}
A high-quality synthesis should maximize semantic coverage rather than concentrate on a narrow theme. We quantify redundancy by computing the average maximum cosine similarity between embeddings of generated question pairs~\cite{chen-etal-2021-training}. Higher redundancy indicates limited analytical breadth, where the model repeatedly focuses on a single topic while overlooking other latent concerns and divergent perspectives. Detailed calculations are provided in Appendix~\ref{subsection:appendix-information}.

%% file: table/data_statistic.tex
\renewcommand{\arraystretch}{1.1}
\begin{table*}[t]
\caption{Comparison with existing related benchmarks. "Real-world" indicates whether the data is sourced from real-world scenarios rather than synthetic or online existing resources. "Live Update" denotes whether the benchmark can be regularly updated.}
    \vspace{-0.1in} 
    \centering
    \small
    \setlength{\tabcolsep}{4pt} 
      \begin{tabular}{lccccc}
        \toprule
        Benchmark & \makecell{Domain} & \makecell{Tasks} & \makecell{Real \\ World} & \makecell{Live \\ Update} \\
        \midrule
        DABstep~\cite{egg2025dabstep} & Data Science & 450 & \cmark & \xmark \\
        FutureX~\cite{zeng2025futurexadvancedlivebenchmark} & Future Prediction & 500/week & \cmark & \cmark \\
        GAIA~\cite{mialon2023gaia} & General QA & 466 & \cmark & \xmark \\
        OSWorld~\cite{xie2024osworld} & Computer Use & 369 & \cmark & \xmark \\
        OPT-Bench~\cite{li2025opt} & Iterative Optimization  & 30 & \cmark & \xmark \\
        Spider2.0~\cite{lei2024spider} & Text-to-SQL & 632 & \cmark & \xmark \\
        SWE-Bench~\cite{jimenez2024swe} & Code  & 2,294 & \cmark & \xmark \\
        \midrule
        SociaBench~\cite{chen2024socialbench} & Social Intent & 6,000 & \xmark & \xmark \\
        URS-bench~\cite{wang2024usercentricmultiintentbenchmarkevaluating} & Intent Understanding & 1,846 & \xmark & \xmark \\
        \bench (ours) & Consumer Intent & 1,475 & \cmark & \cmark \\
        \bottomrule
      \end{tabular}%
    \label{tab:bench}
    \vspace{-1.5em}
\end{table*}

%% file: Fig/pipeline.tex
\begin{figure*}[t]
    \centering
    \includegraphics[width=1\linewidth]{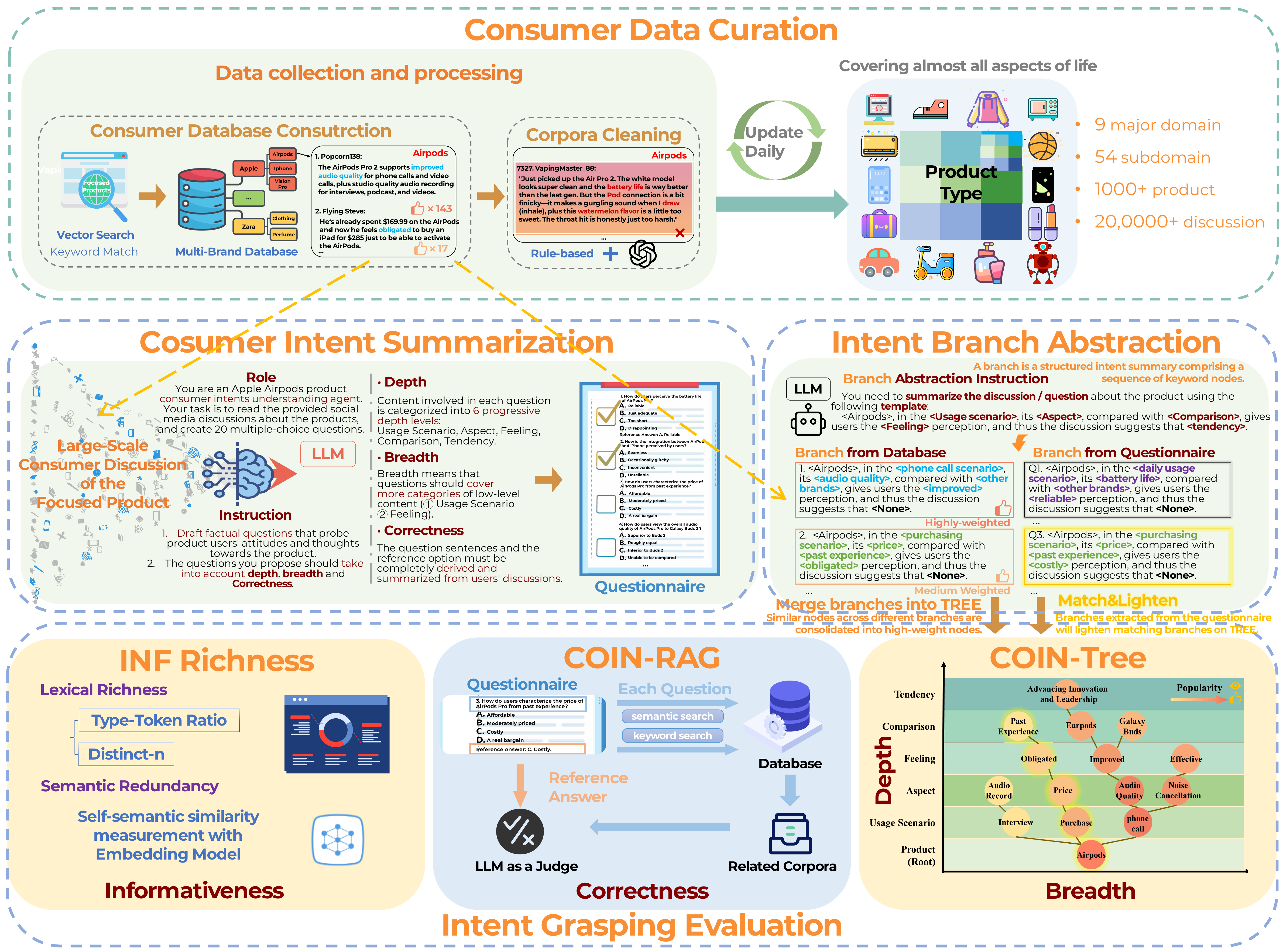}
    \caption{\textbf{\bench Pipeline: From Raw Discourse to Structured Evaluation.}
    \textbf{Top (Curation):} An automated pipeline that consolidates 200k+ public opinions into a consumer database via hybrid keyword–semantic retrieval and dual-stage rule/LLM denoising.
    \textbf{Middle (Active Synthesis):} The \textit{Active Probing} stage, where LLMs act as meta-analysts to convert fragmented discourse into structured questionnaires.
    \textbf{Bottom (Evaluation):} Performance is assessed along four dimensions: \textbf{Depth} and \textbf{Breadth} via \textbf{\textsc{CoIn-Tree}} (a five-level hierarchical intent map), \textbf{Correctness} via \textbf{\textsc{CoIn-RAG}} (grounded corpus verification), and \textbf{Informativeness} (lexical and semantic richness).
    }
    \label{fig:pipeline}
    \vspace{-1.5em}
\end{figure*}

%% file: section/3experiment.tex
\section{Experiment}

\subsection{Experimental Setup}

We evaluated our method on a diverse range of reasoning and general LLMs. This includes proprietary models (GPT family, Claude) accessed via API, and open-source models (Qwen 1.5B–72B, LLaMA, DeepSeek, InternLM) deployed locally via LMDeploy.

\subsection{Main Results}
\label{subsection:main_results}
\input{table/main_results}

Table~\ref{tab:main} presents the evaluation of 20 leading LLMs on \textsc{CoInBench}. Our results reveal a critical performance dichotomy between surface-level information aggregation and deep intent synthesis.

\paragraph{Hierarchy of Intent.}
We observe a pronounced performance decay as we move from explicit to implicit intent layers within \textsc{CoIn-Tree}. While most models maintain reasonable scores in capturing \textit{Usage Scenarios} (L1), their performance collapses when tasked with inferring \textit{Future Tendencies} (L5). This "depth cliff" validates the hierarchical design of our benchmark and highlights a fundamental reasoning bottleneck: models can easily parse what is explicitly stated but struggle to synthesize what is implicitly anticipated by a collective population.

\paragraph{Reasoning vs.\ General Models.}
Reasoning models, particularly \textbf{GPT-o3}, demonstrate a distinct superiority in the deepest cognitive layers (L5) and \textbf{Correctness} (80.35). Unlike general-purpose models that often aggregate noise, \textbf{GPT-o3} exhibits "critical deliberation"—the ability to filter out the vocal minority and anchor its synthesis in true majority consensus. This result suggests that navigating swarm-like discourse requires not just memory, but the ability to perform causal inference and temporal projection to resolve contradictions within the data.

\paragraph{Breadth vs. Precision.}
An intriguing pattern emerges with large-scale open-source models like \textbf{Qwen3-30B-A3B}. While it excels in \textbf{Breadth} (53.20) and surface-level recall (L1), it suffers from high \textbf{Semantic Redundancy} (68.94) and lower \textbf{Correctness}. This indicates a "Verbosity Trap": the model achieves high coverage by exhaustively listing possible viewpoints (brute-force recall) rather than distilling them into a concise, high-entropy consensus. In contrast, \textbf{GPT-5} demonstrates superior semantic span, capturing the "long-tail" of niche concerns while maintaining higher analytical density.

\paragraph{Analytical Density and Information Richness.}
Our \textbf{Informativeness} metrics reveal that \textbf{GPT-o3} achieves the highest lexical richness (85.52) and the lowest semantic redundancy (52.27). This implies that reasoning models actively restructure raw discussion points into a "distilled" analytical form, whereas general models often "parrot" dominant topics, leading to repetitive and diluted outputs. Furthermore, a significant performance gap remains between proprietary models and open-source models, particularly in \textbf{Correctness}, where even the strongest open-source models (e.g., Qwen3-32B) trail behind the GPT-4/5 series by nearly 15-20\%.

\paragraph{Scaling Effects.}
A clear scaling law is observed: models with fewer than 7B parameters (e.g., Qwen2.5-1.5B, DS-7B) fail significantly across all dimensions, particularly in \textbf{Correctness} and \textbf{Breadth}. This underscores that understanding collective intent is an emergent capability that requires a threshold of world knowledge and reasoning capacity to distinguish signal from noise in real-world public discourse.

\subsection{Ablation Study: The Trade-off Between Structure and Noise}
\input{table/abl}



The \tree distills raw discussions into nodes weighted by topic popularity, effectively highlighting mainstream discourse while marginalizing peripheral noise. To investigate the impact of raw swarm noise on model performance, we conducted an ablation study where models were evaluated using the structured \tree as input instead of raw discussions. The results in Table~\ref{tab:abl} reveal a fascinating 
{Structure-Noise Trade-off}:


\noindent\textbf{Structure as a Scaffold for Small Models:} For smaller open-source models (e.g., Qwen2.5-7B), using the \tree significantly improves \textit{Correctness}. Raw noise often overwhelms these noise-intolerant models. The \tree acts as a scaffold that clarifies core intents, revealing both the distribution and logical progression with those intents.

\noindent \textbf{Structure as a Bottleneck for SOTA Models.}
For Proprietary models such as GPT‑4o, applying \tree leads to declines in \textit{Correctness} and \textit{Informativeness}. Although the tree aids denoising, it is inherently \textit{lossy} for SOTA models: while it highlights core intents, it suppresses the subtle nuances they can naturally recover from raw discourse. This suggests that expert‑level intent understanding depends on navigating signal within noise, rather than relying solely on pre‑structured summaries.
    
    

\subsection{Case Study: Quality vs Quantity}
\label{subsection:discussion}

The quantitative results hinted at a discrepancy: \textbf{Qwen3-30B-A3B} outperforms \textbf{GPT-5} at the shallowest level (L1) but lags significantly in deeper reasoning (L2-L4), breadth and informativeness. We further analyze this via a case study on the 'Google Nest' discussion (detailed in Appendix~\ref{subsection:Scores on Nest Smart Speaker}-\ref{subsection:lighted tree}).


A fine-grained qualitative analysis reveals a clear divergence in the \textbf{analytical resolution} of collective intent. Advanced reasoning models such as \textbf{GPT-5} and \textbf{GPT-o3} exhibit strong \textbf{structural inference}, uncovering high-level latent concerns—e.g., \textit{“Matter protocol interoperability”}—that go beyond surface keywords. Rather than merely aggregating terms, these models perform causal mapping to identify the underlying logic of user sentiment.

In contrast, \textbf{Qwen3-30B-A3B} often shows \textbf{semantic dilution}, generating verbose but redundant outputs (e.g., repeated variants of \textit{“Audio Fidelity,”} \textit{“Clarity,”} and \textit{“Sound Quality”}) with limited taxonomic precision. While its MoE architecture supports broad recall, it lacks the \textbf{inferential rigor} for \textbf{discriminative distillation}—i.e., compressing diverse signals into mutually exclusive and collectively exhaustive insights. These findings highlight that understanding collective intent requires not just output breadth but high \textbf{analytical density} and structured synthesis. Expert-level models must demonstrate both depth and breadth in reasoning.



%% file: table/main_results.tex
\begin{table*}[t]
    \caption{Performance of reasoning LLMs, general LLMs, and open-source LLMs on \bench, with the best performance highlighted in \textbf{bold}.}
    \centering
    \small
    \resizebox{\textwidth}{!}{
    \begin{tabular}{lccccccccccc} 
    \toprule
    \multirow{2}{*}{\textbf{Model}} &
      \multicolumn{6}{c}{\textbf{Depth}} & \multirow{2}{*}{\textbf{Breadth}} & \multicolumn{2}{c}{\textbf{Informativeness}} & \multirow{2}{*}{\textbf{Correctness}} \\
    \cmidrule(lr){2-7} \cmidrule(lr){9-10} 
     & 
      \textbf{L1} & \textbf{L2} & \textbf{L3} & \textbf{L4} & \textbf{L5} & \textbf{Overall} & & \textbf{Lexical} & \textbf{Semantic$\downarrow$}  \\
    \midrule
    \rowcolor{lightgray}
    \multicolumn{11}{c}{\emph{Proprietary LLMs}} \\ 
    \midrule
    GPT-5.2 & 23.25 & 25.51 & 6.11 & 3.16 & 0 & \textbf{11.61} & \textbf{53.62} & 76.79 & 65.66 & 85.66  \\  
    Gemini3-Flash & 23.81 & 24.90 & 5.81 & 3.16 & 0.05 & 11.54 &53.51 & 77.65 & 56.62 & \textbf{90.06} \\  
    GPT-5 & 18.29 & \textbf{25.81} & \textbf{6.25} & \textbf{3.49} & 0.06 & 10.78 & 53.48 & 80.21 & 62.75  & 62.65 \\  
    GPT-4.1 & 20.97 & 25.03 & 5.90 & 3.37 & 0 & 11.01 & 53.41 & 79.07  & 63.82  & 59.05 \\ 
    GPT-4o & 20.99 & 23.70 & 4.90 & 2.58 & 0.05 & 10.44 & 52.95 & 79.56  & 62.86  & 75.75 \\ 
    Claude-3.5-sonnet & 19.25 & 23.25 & 5.34 & 2.95 & 0 & 10.16 & 52.83 & 73.94  & 61.11  & 53.35 \\ 
    GPT-o3 & 16.17 & 22.43 & 5.69 & 3.18 & \textbf{0.07} & 9.51 & 52.73 & \textbf{85.52} & \textbf{52.27} & 80.35 \\
    \midrule
    \rowcolor{lightgray}
    \multicolumn{11}{c}{\emph{Open-Source LLMs}} \\ 
    \midrule
    
    Qwen3-30B-A3B & \textbf{25.01} & 23.66 & 5.43 & 2.51 & 0.06 & 11.33 & 53.20 & 70.58 & 68.94 & 61.60 \\
    DS-Distill-Qwen-14B& 17.00 & 25.56 & 6.12 & 3.30 & 0 & 10.40 & 53.32 & 67.47 & 75.53 & 58.45 \\
    Qwen2.5-32B-Instrcut & 19.26& 23.63 & 5.31& 2.89 & 0.01& 10.21& 52.46 & 65.72  & 74.60  &  54.95  \\
    Qwen3-32B & 20.77 & 23.46 & 5.57 & 2.82 & 0 & 10.52 & 51.15 & 65.84 & 68.16 & 55.26 \\
    Qwen3-8B & 15.95 & 22.43 & 4.89 & 2.39 & 0.01 & 9.13 & 50.58 & 57.51 & 	81.25   & 50.42 \\
    Qwen2.5-72B-Instrcut& 18.89 & 22.27  & 5.73 & 3.10 & 0 & 10.00 & 50.52 & 54.63 & 77.49 & 64.11\\ 
    DS-Distill-Qwen-32B& 16.60 & 24.56 & 5.90 & 3.30 & 0 & 10.07 & 50.34 & 59.30 & 76.58 & 53.90 \\
    Qwen2.5-14B-Instrcut & 13.56 & 22.23 & 5.45 & 2.87 & 0.02 & 8.83 & 48.27 & 52.39 & 80.06 & 60.88\\
    LLama3.2-8B-Instrcut & 13.88 & 19.75 & 5.62 & 2.73 & 0 & 8.40 & 47.91 & 47.87  & 88.25  & 52.31 \\  
    Qwen2.5-7B-Instrcut & 11.87 & 19.73 & 4.16 & 1.97 & 0 & 7.54 & 47.43 & 43.58 & 85.07  & 49.24 \\ 
    Internlm3-8B-Instrcut & 11.07 & 20.76 & 4.87 & 2.61 & 0.03 & 7.87 & 45.91 & 49.83 & 75.51  & 51.67 \\
    LLama3.1-8B-Instrcut & 11.23 & 19.46 & 5.53 & 2.91 & 0 & 7.83 & 45.41& 42.36 & 88.00 & 52.67 \\
    Qwen2.5-3B-Instrcut & 13.49 & 18.63 & 4.22 & 2.09 & 0 & 7.69 & 42.73 & 39.32 & 79.35  & 35.43 \\ 
    Qwen2.5-1.5B-Instrcut &2.83 & 4.94 & 0.99 & 0.45 & 0 & 1.84 & 14.31& 4.56 & 87.65 & 36.90\\
    DS-Distill-Qwen-7B& 1.80 & 4.91 & 1.35 & 0.55 & 0 & 1.72 & 11.54 & 3.30 & 73.25 & 13.40 \\
    \bottomrule
    \end{tabular}
    }
    \label{tab:main}
    \label{-3em}
\end{table*}

%% file: table/abl.tex
\begin{table*}[t]
    \caption{Comparison of reasoning LLMs, general LLMs, and open-source LLMs using \tree on \bench, with the best performance highlighted in \textbf{bold}.}
    \centering
    \small
    \resizebox{\textwidth}{!}{
    \begin{tabular}{lccccccccccc} 
    \toprule
    \multirow{2}{*}{\textbf{Model}} &
      \multicolumn{6}{c}{\textbf{Depth}} & \multirow{2}{*}{\textbf{Breadth}} & \multicolumn{2}{c}{\textbf{Informativeness}} & \multirow{2}{*}{\textbf{Correctness}} \\
    \cmidrule(lr){2-7} \cmidrule(lr){9-10} 
     & 
      \textbf{L1} & \textbf{L2} & \textbf{L3} & \textbf{L4} & \textbf{L5} & \textbf{Overall} & & \textbf{Lexical} & \textbf{Semantic$\downarrow$}  \\
       \midrule
     GPT-5.2 (wo/Tree) & 23.25 & 25.51 & 6.11 & 3.16 & 0 & 11.61 & 53.62 & 76.79 & 65.66 & 85.66\\
     GPT-5.2 (w/Tree)   & 44.71 & 34.11 & 13.94 & 10.51 & 0.46 & 20.75 & 58.37 & 68.71  & 73.71 & 56.35  \\
     & (\textcolor{red}{+21.46}) & (\textcolor{red}{+8.60}) & (\textcolor{red}{+7.83}) & (\textcolor{red}{+7.35}) & (\textcolor{red}{+0.46}) & (\textcolor{red}{+9.14}) & (\textcolor{red}{+4.75}) & (\textcolor{green}{-8.08}) & (\textcolor{green}{-8.05}) & (\textcolor{green}{-29.31}) \\
     \midrule
     Gemini3-Flash (wo/Tree)& 23.81 & 24.90 & 5.81 & 3.16 & 0.05 & 11.54 &53.51 & 77.65 & 56.62 &  90.06\\
     Gemini3-Flash (w/Tree)   & 41.32 & 33.97 & \textbf{15.24} & \textbf{11.74} & \textbf{1.95} & 20.84 & 58.63 & 68.39  & \textbf{68.55} &  \textbf{71.65}\\
     & (\textcolor{red}{+17.51}) & (\textcolor{red}{+9.07}) & (\textcolor{red}{+9.43}) & (\textcolor{red}{+8.58}) & (\textcolor{red}{+1.90}) & (\textcolor{red}{+9.30}) & (\textcolor{red}{+5.12}) & (\textcolor{green}{-9.26}) & (\textcolor{green}{-8.07}) & (\textcolor{green}{-18.31}) \\
    
    \midrule
    GPT-o3 (wo/Tree) & 16.17 & 22.43 & 5.69 & 3.18 & 0.07 & 9.51 & 52.73 & 85.52  & 52.27  & 80.35 \\
    GPT-o3 (w/Tree)   & \textbf{45.13} & \textbf{36} & 15 & \textbf{11.74} & 1.37 & \textbf{21.95} & \textbf{59.16} & \textbf{72.47}  & 71.05 & 57.10 \\
    & (\textcolor{red}{+28.96}) & (\textcolor{red}{+13.57}) & (\textcolor{red}{+9.31}) & (\textcolor{red}{+8.56}) & (\textcolor{red}{+1.30}) & (\textcolor{red}{+12.44}) & (\textcolor{red}{+6.43}) & (\textcolor{green}{-13.05}) & (\textcolor{green}{-0.22}) & (\textcolor{green}{-23.25}) \\
    \midrule
    GPT-4o (wo/Tree)& 20.99 & 23.70 & 4.90 & 2.58 & 0.05 & 10.44 & 52.95 & 79.56  & 62.86  & 75.60 \\
    GPT-4o (w/Tree) & 37.38 & 31.39 & 12.78 & 9.60 & 0.85 & 18.40 & 57.00 & 70.77 & 72.36 & 64.15 \\
    & (\textcolor{red}{+16.39}) & (\textcolor{red}{+7.69}) & (\textcolor{red}{+7.88}) & (\textcolor{red}{+7.02}) & (\textcolor{red}{+0.80}) & (\textcolor{red}{+7.96}) & (\textcolor{red}{+4.05}) & (\textcolor{green}{-8.79}) & (\textcolor{green}{-0.50}) & (\textcolor{green}{-11.45}) \\
    \midrule
    Qwen2.5-7B (wo/Tree) & 11.87 & 19.73 & 4.16 & 1.97 & 0 & 7.54 & 47.43 & 43.58  & 85.07 & 42.15 \\ 
    Qwen2.5-7B (w/Tree) & 32.39 & 29.78 & 11.29 & 7.96 & 0.54 & 16.39 & 49.28  & 42.33  & 79.63 & 49.24 \\
    & (\textcolor{red}{+20.52}) & (\textcolor{red}{+10.05}) & (\textcolor{red}{+7.13}) & (\textcolor{red}{+6.00}) & (\textcolor{red}{+0.54}) & (\textcolor{red}{+8.85}) & (\textcolor{red}{+1.85}) & (\textcolor{green}{-1.25}) & (\textcolor{green}{-5.44}) & (\textcolor{red}{+7.09}) \\
    \bottomrule
    \end{tabular}
    }
    \label{tab:abl}
    \vspace{-1.5em}
\end{table*}

 \begin{table*}[t]
     \caption{30 Expert Questionnaires.}
     \centering
     \small
     \resizebox{\textwidth}{!}{
     \begin{tabular}{lccccccccccc} 
     \toprule
     \multirow{2}{*}{\textbf{Source}} &
       \multicolumn{6}{c}{\textbf{Depth}} & \multirow{2}{*}{\textbf{Breadth}} & \multicolumn{2}{c}{\textbf{Informativeness}} & \multirow{2}{*}{\textbf{Correctness}} \\
     \cmidrule(lr){2-7} \cmidrule(lr){9-10} 
      & 
       \textbf{L1} & \textbf{L2} & \textbf{L3} & \textbf{L4} & \textbf{L5} & \textbf{Overall} & & \textbf{Lexical} & \textbf{Semantic$\downarrow$}  \\
        \midrule
     
     GPT-5.2 & 23.76 & 30.66 & 7.67 & 4.13 & 0 & 13.24 & 55.32 & 79.47   & 63.33  & 86.67  \\
     Expert & 46.42 & 37.95 & 15.80 & 12.27 & 1.77 & 22.84 & 60.13 &73.07  &67.65  & 88.17 \\
     & (\textcolor{red}{+22.66}) & (\textcolor{red}{+7.29}) & (\textcolor{red}{+8.13}) & (\textcolor{red}{+8.14}) & (\textcolor{red}{+1.77}) & (\textcolor{red}{+9.60}) & (\textcolor{red}{+4.81}) & (\textcolor{green}{-6.40}) & (\textcolor{green}{-4.32}) & (\textcolor{red}{+1.50}) \\

    GPT-DeepResearch & 18.25 & 26.79 & 7.23 & 4.63 & 0 & 11.37 & 54.56 &79.88  &58.55  & 71.60  \\

     \bottomrule
     \end{tabular}
     }
     \label{tab:expert}
     \vspace{-1.5em}
 \end{table*}

%% file: section/2related_work.tex
\section{Related Work}
\noindent\textbf{LLM Evaluation}  
The rapid progress of Large Language Models (LLMs) has led to a proliferation of benchmarks for evaluating their generalization and reasoning abilities. Early benchmarks such as MMLU~\cite{hendrycks2020measuring} and BIG-bench~\cite{srivastava2022beyond} assessed broad knowledge and reasoning. Later efforts targeted specific domains, including linguistic and commonsense reasoning~\cite{wang2018glue, wang2019superglue, talmor2019commonsenseqa, zellers2019hellaswag, lin2022truthfulqa}, mathematical and programming tasks~\cite{hendrycks2021measuring, cobbe2021training, chen2021evaluating, austin2021program}, and agent-based evaluations~\cite{chan2024mle, xie2024osworld, li2025np, li2025opt}. 
Despite these advances, existing benchmarks rarely examine whether LLMs can understand human intent in dynamic, real-world decision-making—particularly in multi-user, emotionally nuanced, and evolving contexts. To address this, we propose \bench, a large-scale benchmark designed to assess LLMs’ ability to comprehend and reason about human intentions in complex real-world scenarios.

\noindent\textbf{LLM Human Intent Evaluation}  
Human intent evaluation has increasingly shifted toward understanding human‑centric intent in complex, real‑world contexts. Prior benchmarks capture different facets of this problem: SocialIQA~\cite{sap2019socialiqa} focuses on social and commonsense intent, TOMI~\cite{le2019revisiting} evaluates Theory of Mind capabilities, and IFEVAL~\cite{zhou2023instruction} measures instruction‑following. Other efforts such as SociaBench~\cite{chen2024socialbench}, AgentSense~\cite{mou2025agentsense}, EmotionQueen~\cite{chen2024emotionqueenbenchmarkevaluatingempathy}, and URS‑bench~\cite{wang2024usercentricmultiintentbenchmarkevaluating} examine intent understanding from social, emotional, or user‑response perspectives.
However, existing frameworks typically isolate single dimensions of intent, whereas real‑world intent is multifaceted and dynamic, combining social, emotional, and practical factors. Consequently, there remains no comprehensive benchmark for evaluating whether LLMs can fully understand human reasoning and mental states. To address this gap, we propose \bench, a benchmark for assessing LLMs’ ability to understand complex, dynamic real‑world human intent.

%% file: section/4conclusion.tex
\vspace{-5pt}
\section{Conclusion}
\vspace{-5pt}
In this work, we propose \bench, a comprehensive benchmark consisting of over 200k product-level discussions across 9 major domains, 54 sub-domains, and over 1,400 products, designed to evaluate the performance of Large Language Models (LLMs) in understanding real-world human intent, particularly within consumer domains. Our evaluation framework measures LLMs' ability to comprehend intent across four key dimensions: depth, breadth, correctness, and informativeness. We implement a robust evaluation pipeline to mitigate bias and hallucinations. Specifically, we construct \tree to assess LLMs' depth and breadth of intent understanding, use \rag for evaluating correctness, and measure informativeness through lexical diversity and semantic richness. 
Through extensive experiments on both closed-source and open-source models, we demonstrate that reasoning models outperform general models on average. However, significant gaps remain between closed-source and open-source models, and even the most advanced models struggle with deep and broad intent understanding. Our mission is to advance LLMs toward expert-level reasoning and improve their ability to understand complex real-world intent.

%% file: section/appendix.tex
\newpage
\section{Appendix}

\subsection{Use of Large Language Models}
Large Language Models are used for grammar check and polishing in this paper.
\subsection{Data Statement}
The \bench dataset was constructed from public available websites, and all privacy-sensitive personal information has been removed during the data curation process. To mitigate the potential for technology misuse, the benchmark will be released under a restrictive license for academic research purposes only, explicitly prohibiting malicious applications.
\subsection{Data Statistics}
\label{apd:data_stat}
\input{Fig/data}
In this section, we provide a detailed statistical breakdown of the \textsc{\bench} dataset. To ensure a comprehensive evaluation of collective intent across diverse human activities, we constructed a two-tier hierarchical taxonomy consisting of 9 primary domains and 54 fine-grained sub-categories, as visualized in figure~\ref{fig:data}.

\paragraph{Domain Distribution and Scale}
As illustrated in the inner ring of our taxonomy, \textsc{\bench} covers the breadth of the modern consumer ecosystem. The dataset is anchored by major sectors including \textit{Home and Daily Necessities} (284 products), \textit{Consumer Electronics} (201 products), \textit{Software} (199 products), and \textit{Vehicles} (166 products). Other significant domains include \textit{Personal Care and Beauty} (145), \textit{Fashion and Accessories} (142), \textit{AI Field} (125), \textit{Sports and Fitness} (111), and \textit{Food and Beverage} (102). This distribution ensures that LLMs are tested not only on general consumer goods but also on high-context, specialized fields such as AI technologies and industrial software.

\paragraph{Granular Sub-categories}
The outer ring of the sunburst chart highlights the high granularity of our benchmark. Each primary domain is further decomposed into specific sub-categories to capture nuanced usage scenarios and collective intents. For instance:
\begin{itemize}
    \item \textbf{High-Tech \& Digital:} Includes niche sectors like \textit{AI Learning Machines}, \textit{NLP/CV} tools, and \textit{Programming/Development Tools} (the largest sub-category with 30 products).
    \item \textit{Lifestyle \& Essentials:} Spans from \textit{Skin Care Products} and \textit{Individual Care} to \textit{Kitchen Appliances} and \textit{Fresh Food}.
    \item \textbf{Mobility:} Covers the transition from traditional \textit{Fuel Vehicles} to \textit{Electric Vehicles} and \textit{New Means of Transportation}.
\end{itemize}

\paragraph{Data Consistency and Density}
A key feature of \textsc{\bench} is its balanced distributional design. Most sub-categories are populated with approximately 28 to 29 distinct product discussions, ensuring that the benchmark is not skewed toward any single niche. With an average of 200 user comments per product, the total corpus comprises over 200,000 unique human perspectives. This density provides a sufficiently "noisy" and "swarm-like" environment, forcing models to demonstrate advanced synthesis and denoising capabilities rather than simple pattern matching.

\subsection{\tree construction details}
\label{subsection: tree}

First, the LLM (GPT-4o) is utilized to extract branches from each consumer discussion. During this process, the model is prompted to summarize a user’s discussion following the template: \textlangle {{Product\_series}}\textrangle, in the \textlangle Usage Scenario\textrangle, its \textlangle Aspect\textrangle, compared with \textlangle Comparison\textrangle, gives consumers the \textlangle Feeling\textrangle~perception, and the discussion suggests that \textlangle Tendency\textrangle.

Next, the branches are used to construct the tree. All branches are connected to the tree root node, forming an initial tree. Sentence Transformers are then employed to merge semantically similar nodes layer by layer from the top down within the initial tree. During this process, nodes in the same layer that share the same parent node and are semantically similar are merged into one. The child nodes of each pre-merged node are then designated as the child nodes of the merged node. Additionally, the weight of the merged node is calculated as the sum of the weights of its child nodes. Ultimately, the fully merged tree is referred to as \tree, which will resemble the structure shown in figure X. Nodes with higher weights will appear in the shallow layers of the tree; this is because the core topics of discussion (e.g., usage scenarios, aspects, feelings) often overlap across discussions from different consumers, and such high-weight nodes represent the aspects of the product that users focus on most.

This process enables the clear presentation of user discussion content in a tree structure while highlighting discussion hotspots. Meanwhile, by updating the discussion data and reconstructing \tree, we can analyze changes in child nodes under the same parent node between the two trees—thereby identifying users’ immediate concerns and long-term strategic considerations.

After summarization, the key terms in the template are fetched to form a single branch. For the branches derived from one discussion, the initial weight of each node is equal, ranging from 1 to 3, and determined by the discussion’s upvotes and view count. Notably, not every discussion can be summarized to fill all six key terms—more successfully filled key terms correspond to a longer branch path, which in turn reflects a more in-depth consumer intent.
Next, all branches are connected to the tree root node, forming an initial tree. Sentence Transformers are then employed to merge semantically similar nodes layer by layer from the top down. During this process, nodes in the same layer that share the same parent node and are semantically similar are merged into one. The child nodes of each pre-merged node are then designated as the child nodes of the merged node. Additionally, the weight of the merged node is calculated as the sum of the weights of its child nodes. Ultimately, the fully merged tree is referred to as \tree. Nodes with higher weights will appear in the shallow layers of the tree; this is because the core topics of discussion (e.g., usage scenarios, aspects, feelings) often overlap across discussions from different consumers.

\noindent\textbf{Lighten the Tree:} To assess the depth and breadth of a LLM’s understanding of consumer intent, content from questionnaires will be extracted into branches. Each branch will undergo semantic matching with the nodes in the \tree from top to bottom using a Sentence Transformer. Nodes that are successfully matched will be marked as "lightened," and the lightened nodes in the \tree will form a subtree. and the questionnaire will receive the score corresponding to that node.

Specifically, for each question in the questionnaire, the question stem and its four options will first be concatenated into four opinion statements. The LLM will then extract four branches from these four statements. These four branches will be matched with the nodes in the \tree from top to bottom—each branch will lighten a path and obtain a score based on the weights of the nodes along that path. The branch with the highest score for a given question will be used to "lighten" the \tree. Notably, nodes in the \tree cannot be repeatedly lightened by different questions. After iterating through all questions, the lightened nodes in the \tree will form a subtree.

\subsection{Informativeness}
\label{subsection:appendix-information}
\noindent\textbf{Lexical Richness:} The evaluation of lexical richness relies on two key metrics: Type-Token Ratio (TTR) \cite{johnson1944type} and Distinct-n \cite{li2016diversity}. TTR quantifies the ratio of unique tokens to the total number of words in the text. It is defined as:
\begin{equation}\label{TTR}
\begin{aligned}
\notag
\text{TTR} = \frac{\text{Count(unique token)}}{\text{Count(tokens)}}
\end{aligned}
\end{equation}
where a higher TTR indicates greater lexical richness. Distinct-n focuses on the \textit{n-gram} level, measuring the ratio of unique \textit{n-grams} to the total number of \textit{n-grams}. This study focuses on \textit{bi-grams}, and the Distinct-n is calculated as:
\begin{equation}\label{distint-n}
\begin{aligned}
\notag
\text{Distinct-n} = \frac{\text{Count(unique bi-gram)}}{\text{Count(bi-grams)}}.
\end{aligned}
\end{equation}
\textbf{Semantic Redundancy} is evaluated using a self-referential manner \cite{chen-etal-2021-training}, where 
the average maximum semantic similarity is computed between each question and all other questions in the questionnaire, as well as between each question’s options and all other questions' options. Given a set of questions \( Q = \{ q_1, q_2, ..., q_n \} \), the semantic similarity between any two questions \( q_i \) and \( q_j \) is calculated using cosine similarity:

\begin{equation}\label{S}
\begin{aligned}
\notag
\text{Sim}(q_i, q_j) = \frac{\mathbf{v_i} \cdot \mathbf{v_j}}{\|\mathbf{v_i}\| \|\mathbf{v_j}\|},
\end{aligned}
\end{equation}

where \(\mathbf{v_i}\) and \(\mathbf{v_j}\) represent the vector embeddings of questions \(q_i\) and \(q_j\), respectively. The redundancy score is then computed as the average of the maximum similarity values across all pairs of questions:

\begin{equation}\label{Redundancy}
\begin{aligned}
\notag
\text{Redundancy} = \frac{1}{n} \sum_{i=1}^{n} \max_{j \neq i} \text{Sim}(q_i, q_j).
\end{aligned}
\end{equation}

Notably, a lower redundancy score indicates less repetition in question paradigms and option designs, which reflects the LLM’s ability to understand consumers’ intentions from multiple perspectives and conduct multi-source causal inference.

\subsection{Case Study}
\label{subsection:Scores on Nest Smart Speaker}

\begin{table*}[h]
    \caption{Performance of reasoning LLMs, general LLMs, and open-source LLM on Google Nest Smart Discussion.}
    \centering
    \small
    \resizebox{\textwidth}{!}{
    \begin{tabular}{lccccccccccc} 
    \toprule
    \multirow{2}{*}{\textbf{Model}} &
      \multicolumn{6}{c}{\textbf{Depth}} & \multirow{2}{*}{\textbf{Breadth}} & \multicolumn{2}{c}{\textbf{Informativeness}} & \multirow{2}{*}{\textbf{Correctness}} \\
    \cmidrule(lr){2-7} \cmidrule(lr){9-10} 
     & 
      \textbf{L1} & \textbf{L2} & \textbf{L3} & \textbf{L4} & \textbf{L5} & \textbf{Overall} & & \textbf{Lexical} & \textbf{Semantic$\downarrow$}  \\
    \midrule
    GPT-5 & 3.33 & \textbf{21.79} & \textbf{6.53} & \textbf{4.19} & 0.00 & 7.17 & \textbf{50.84} & \textbf{0.87} & \textbf{0.30} & 0.75\\
    GPT-o3 & 2.51 & 20.02 & 5.33 & 2.77 & \textbf{1.91} & 6.51 & 50.19 & 0.75 & 0.46 & \textbf{0.95}\\
    Qwen3-30B-A3B &\textbf{31.32} & 16.61 & 2.77 & 1.49 & 0.00 & \textbf{10.44} &	50.51 &	0.81 & 0.39 & 0.75 \\

    \bottomrule
    \end{tabular}
    }
    \label{tab:Google Nest Smart Speaker}
\end{table*}

\subsubsection{Google Nest Smart Speaker questionnaire from GPT-5}
\label{subsection:GPT-5 questionnaire}
\begin{enumerate}

\item \textbf{How do you primarily use your Google Nest speakers at home?}
\begin{enumerate}[label=\Alph*.]
    \item For music playback
    \item For controlling smart devices
    \item For asking questions/time/weather
    \item For security alerts or doorbell announcements
\end{enumerate}
\textit{Answer: A. Users reported using Nest speakers most often for music, followed by smart home control and daily reminders such as weather or timers.}

\item \textbf{How satisfied are you with the sound quality of Nest Audio compared to Nest Mini?}
\begin{enumerate}[label=\Alph*.]
    \item Nest Audio is leagues better, especially bass
    \item Mini is enough for casual listening
    \item Nest Audio is adequate but not impressive
    \item No difference noticed
\end{enumerate}
\textit{Answer: A. Users consistently said Nest Audio has much better bass and overall quality compared to Mini, making it preferable for music.}

\item \textbf{Have you experienced connection issues with your Nest speakers in recent years?}
\begin{enumerate}[label=\Alph*.]
    \item Yes, frequent disconnections and 'sorry something went wrong'
    \item Yes, occasional hiccups
    \item No, they work reliably
    \item Issues only due to Wi-Fi provider/router
\end{enumerate}
\textit{Answer: A. Many users reported worsening connection reliability over time, though some fixed issues by upgrading Wi-Fi or resetting devices.}

\item \textbf{How well does Google Home integrate with your non-Google devices (like Tuya, ZigBee, or Ikea smart products)?}
\begin{enumerate}[label=\Alph*.]
    \item Very smooth integration
    \item Works but often buggy
    \item I cannot integrate them at all
    \item I only use 100\% Google products
\end{enumerate}
\textit{Answer: B. Several users noted persistent issues integrating Tuya/Lidl ZigBee and Ikea products with Google Home compared to their native apps.}

\item \textbf{Which competing ecosystem do you feel has smoother smart home integration?}
\begin{enumerate}[label=\Alph*.]
    \item Apple HomeKit/HomePod
    \item Google Nest/Home
    \item Amazon Alexa/Echo
    \item None of them are ideal
\end{enumerate}
\textit{Answer: A. Some users with Apple devices felt HomePod integration was smoother, while Google or Amazon were preferred for value and compatibility.}

\item \textbf{What’s your biggest frustration with Nest smart speakers?}
\begin{enumerate}[label=\Alph*.]
    \item Laggy responses
    \item Misunderstood commands
    \item Device going offline
    \item Limited software updates
\end{enumerate}
\textit{Answer: A. Users most commonly complained about slow responses and frequent 'didn't understand' errors, especially in recent years.}

\item \textbf{How often do you use your Nest devices for multi-room audio?}
\begin{enumerate}[label=\Alph*.]
    \item Daily
    \item Occasionally for parties/events
    \item Rarely
    \item Never
\end{enumerate}
\textit{Answer: A. Users said multi-room audio used to work seamlessly but has become frustratingly unreliable, with groups often not functioning.}

\item \textbf{Would you consider switching away from Google due to reliability issues?}
\begin{enumerate}[label=\Alph*.]
    \item Yes, to Alexa
    \item Yes, to Apple
    \item Yes, to Home Assistant or local-first systems
    \item No, I will stay with Google
\end{enumerate}
\textit{Answer: A. Some users said they were ready to move to Alexa or Home Assistant because of lag and reliability problems with Nest.}

\item \textbf{Do you find Nest speakers useful for families with children?}
\begin{enumerate}[label=\Alph*.]
    \item Yes, great for music and educational features
    \item Somewhat, but limited reliability frustrates kids
    \item Not useful at all for kids
    \item Only good as background music
\end{enumerate}
\textit{Answer: B. Users mentioned features like 'animal of the day' were fun for kids, but white noise and music playback have become unreliable.}

\item \textbf{How important is price compared to reliability when choosing between Google, Amazon, or Apple smart speakers?}
\begin{enumerate}[label=\Alph*.]
    \item Price is the most important
    \item Balance between price and reliability
    \item Reliability is the most important
    \item Ecosystem integration matters most
\end{enumerate}
\textit{Answer: A. Users often noted that Google's affordability got them locked in, but reliability issues make them consider pricier alternatives.}

\item \textbf{Have you faced limitations when casting media from phones to Nest devices?}
\begin{enumerate}[label=\Alph*.]
    \item Yes, apps don’t always connect
    \item Sometimes there’s lag/delay
    \item No issues at all
    \item I don’t use casting
\end{enumerate}
\textit{Answer: A. Users reported frequent casting issues, especially with Spotify, YouTube, and video streams not reaching certain Nest devices.}

\item \textbf{If you primarily wanted a device for music, which would you pick?}
\begin{enumerate}[label=\Alph*.]
    \item Google Nest Audio
    \item Amazon Echo 4th gen
    \item Apple HomePod Mini
    \item Traditional Bluetooth speakers
\end{enumerate}
\textit{Answer: A. Users debated between Nest Audio, Echo, and HomePod. Many said Nest Audio had good bass but Echo was decent, while some still preferred Sonos or passive Bluetooth sets.}

\item \textbf{How do you feel about Google discontinuing/reducing stock of Nest devices?}
\begin{enumerate}[label=\Alph*.]
    \item Concerned about product support
    \item Neutral, waiting for new models
    \item Considering switching to another brand
    \item Not worried at all
\end{enumerate}
\textit{Answer: A. Several users worried Nest Mini and Audio are discontinued, wondering if Google will abandon the smart speaker hardware.}

\item \textbf{What feature would make you more likely to stick with Google Nest speakers?}
\begin{enumerate}[label=\Alph*.]
    \item Improved reliability and faster response
    \item Better music/audio quality
    \item Deeper integration with third-party devices
    \item Clear roadmap and updates from Google
\end{enumerate}
\textit{Answer: A. Users said better reliability, sound improvements, and smoother ecosystem updates would convince them to remain loyal.}

\item \textbf{Do you experience more issues with Google Assistant understanding you in multilingual households?}
\begin{enumerate}[label=\Alph*.]
    \item Yes, it constantly misinterprets
    \item Sometimes, especially switching languages
    \item No, it works fine in multiple languages
    \item I only use one language
\end{enumerate}
\textit{Answer: A. Users noted Assistant struggles badly in multilingual homes, often failing basic commands or mixing languages incorrectly.}

\item \textbf{What’s your perspective on Nest speakers’ long-term durability?}
\begin{enumerate}[label=\Alph*.]
    \item Still working fine years later
    \item Performance has worsened over time
    \item Hardware is durable but software declines
    \item They feel like e-waste now
\end{enumerate}
\textit{Answer: C. Some users praised durability, while many complained hardware outlasted software support, calling devices obsolete early.}

\item \textbf{How do you primarily resolve issues with Nest devices?}
\begin{enumerate}[label=\Alph*.]
    \item Factory reset
    \item Router and Wi-Fi upgrades
    \item Reinstalling Google Home app
    \item Contacting Google support
\end{enumerate}
\textit{Answer: A. Most users resorted to factory resets or Wi-Fi upgrades; official support was rarely mentioned as helpful.}

\item \textbf{Would you invest in another Nest smart display (like Hub/Hub Max) now?}
\begin{enumerate}[label=\Alph*.]
    \item Yes, I still trust Google ecosystem
    \item Maybe, if I find a second-hand deal
    \item No, too much risk of discontinued support
    \item I prefer other brands’ smart displays
\end{enumerate}
\textit{Answer: C. Users were hesitant to buy discontinued Nest Hubs/Max, fearing bricking or lack of updates.}

\item \textbf{When connecting Nest with services like Spotify or YouTube Music, what’s your experience?}
\begin{enumerate}[label=\Alph*.]
    \item Smooth, works well
    \item Works but occasionally lags
    \item Often breaks or blocks premium-only features
    \item I don’t link music services
\end{enumerate}
\textit{Answer: C. Several users reported Spotify on Nest sometimes says 'premium only' even with premium, and YouTube Music integration often fails.}

\item \textbf{What future direction should Google take with Nest smart speakers?}
\begin{enumerate}[label=\Alph*.]
    \item Bring Gemini AI with better natural understanding
    \item Focus on keeping devices reliable
    \item Produce new affordable hardware
    \item Open-source support if retiring devices
\end{enumerate}
\textit{Answer: B. Users speculated Google must fix reliability, offer Gemini AI improvements, and either release new hardware or open source old devices.}

\end{enumerate}
\subsubsection{Google Nest Smart Speaker questionnaire from GPT-o3}
\label{subsection:GPT-o3 questionnaire}

\begin{enumerate}

\item \textbf{When users retrofit 1980s intercoms with Nest Mini units, which room-specific control do they hope to achieve later on?}
\begin{enumerate}[label=\Alph*.]
    \item Only ceiling fans and lights of that bedroom
    \item Satellite TV channels in the garage
    \item Printer queues in the study
    \item Irrigation valves in the backyard
\end{enumerate}
\textit{Answer: A. Users describe planning “a speaker in every bedroom with some intricate setup to both only control devices specific to that room (like ceiling fans and lights) as well as shared devices.”}

\item \textbf{What adjective did a long-time owner use to praise Nest Audio’s stereo sound after pairing two units?}
\begin{enumerate}[label=\Alph*.]
    \item Incredible
    \item Tinny
    \item Muffled
    \item Overpriced
\end{enumerate}
\textit{Answer: A. A commenter said “I use 2 Nest Audios in a stereo setup, and the audio is incredible,” reflecting positive feelings about sound quality.}

\item \textbf{Which competing smart speaker line did several Redditors say they might switch to because Google devices have become “laggy” and “driving me insane”?}
\begin{enumerate}[label=\Alph*.]
    \item Amazon Echo / Alexa
    \item Sonos Era
    \item Bose Smart Ultra
    \item Marshall Uxbridge
\end{enumerate}
\textit{Answer: A. Many posts mention considering Amazon Echo or Alexa displays as an alternative when Nest performance deteriorated.}

\item \textbf{In the thread about buying a Nest Hub Max, which security-related use case was highlighted as a reason to still want the display?}
\begin{enumerate}[label=\Alph*.]
    \item Acting as a digital photo frame with camera recording
    \item Hosting a VPN server
    \item Controlling sprinklers via Zigbee
    \item Calibrating 3D printers
\end{enumerate}
\textit{Answer: A. A buyer said they liked “the camera/security recording function and using it as a digital photo frame,” showing the usage scenario.}

\item \textbf{Which phrase did frustrated owners repeatedly hear instead of successful commands, prompting them to call Google Home a “support group”?}
\begin{enumerate}[label=\Alph*.]
    \item “Sorry, something went wrong, try again later.”
    \item “Firmware upgrade in progress.”
    \item “Device is paired in another room.”
    \item “Low battery, shutting down.”
\end{enumerate}
\textit{Answer: A. Multiple users quote the device replying “Sorry, something went wrong, try again later,” illustrating a common pain point.}

\item \textbf{Why did one user say the Pixel Tablet on its dock feels like an “old TV/VCR combo” compared with a real Nest Hub?}
\begin{enumerate}[label=\Alph*.]
    \item It can’t be asked to play music on other Google speakers
    \item It lacks Wi-Fi 6E support
    \item The screen is smaller than 5 inches
    \item It forces Amazon Prime ads
\end{enumerate}
\textit{Answer: A. They complained that you “can’t tell it to play music on it from another Google speaker,” so the hybrid device does neither role well.}

\item \textbf{Which connectivity problem did a border-area listener report when TuneIn stations kept dropping on Nest speakers?}
\begin{enumerate}[label=\Alph*.]
    \item Occasional to frequent loss in signal
    \item Crackling Bluetooth interference only at night
    \item Wrong language playback
    \item Overheating power adapters
\end{enumerate}
\textit{Answer: A. The post says “I have experienced occasional to frequent loss in signal when listening to stations that utilize TuneIn.”}

\item \textbf{When discussing Matter devices going offline, which brand of mesh router system was singled out for Thread settings confusion?}
\begin{enumerate}[label=\Alph*.]
    \item Eero 6E
    \item TP-Link Deco
    \item UniFi Dream Router
    \item Netgear Orbi
\end{enumerate}
\textit{Answer: A. A user wrote “I have an Eero 6e mesh router system… The Threads feature is toggled on,” yet their Matter gear still dropped.}

\item \textbf{How did a Nest Mini owner describe the music delay when the speaker was added to a stereo link in Google Home?}
\begin{enumerate}[label=\Alph*.]
    \item The delay is HUGE.
    \item It syncs perfectly.
    \item Only milliseconds of lag.
    \item Delay happens once a month.
\end{enumerate}
\textit{Answer: A. The post states, “If I play any music…the delay is HUGE,” emphasizing a negative feeling about latency.}

\item \textbf{Which future-oriented speculation did shoppers raise after noticing no Nest Audio stock in multiple country stores?}
\begin{enumerate}[label=\Alph*.]
    \item A new generation might be announced at the Pixel event
    \item Google is switching to Apple HomeKit
    \item All smart speakers will become subscription-based
    \item Wi-Fi will be removed from Nest
\end{enumerate}
\textit{Answer: A. They asked, “Are people expecting a new generation to be announced at the Pixel event in a couple weeks?”—a tendency toward anticipating new hardware.}

\item \textbf{Which cloud storage dilemma did dual-ecosystem users discuss while already owning many Nest Hubs and iCloud devices?}
\begin{enumerate}[label=\Alph*.]
    \item Paying for both 200 GB iCloud and 200 GB Google One plans
    \item Choosing between Dropbox and Box free tiers
    \item Losing access to Microsoft OneDrive photos
    \item Migrating from Amazon S3 Glacier Vaults
\end{enumerate}
\textit{Answer: A. The repeated post describes both iCloud and Google One hitting the 200 GB limit and not wanting to upgrade both.}

\item \textbf{What network feature on apartment Wi-Fi prevented an elderly resident’s Nest Mini from completing setup?}
\begin{enumerate}[label=\Alph*.]
    \item AP Isolation turned on
    \item Hidden SSID broadcast
    \item WPA3-Enterprise only
    \item Dual NAT tunneling
\end{enumerate}
\textit{Answer: A. The care home enables “AP Isolation,” so the speaker throws the message “Please check your Wi-Fi network settings.”}

\item \textbf{Which sound-related improvement motivated users to prefer Nest Audio over their old Google Home Minis?}
\begin{enumerate}[label=\Alph*.]
    \item ‘Bass is the most noticeable improvement’ at high volume
    \item Built-in CD player support
    \item Dolby Atmos rear channels
    \item Quad-mic noise cancelling
\end{enumerate}
\textit{Answer: A. One review says, “Bass is the most noticeable improvement, high volume performance is better,” highlighting the aspect of audio quality.}

\item \textbf{How much did Canadian bargain hunters report paying at Lowe’s or Home Depot for clearance Nest Audio units?}
\begin{enumerate}[label=\Alph*.]
    \item \$39.97
    \item \$129.99
    \item \$199.00
    \item \$15.00
\end{enumerate}
\textit{Answer: A. Posts note “Nest Audio for sale for \$39.97… is it worth getting,” reflecting pricing sentiment.}

\item \textbf{Which workaround did some owners adopt because the Nest Hub could no longer resume music on the intended room speaker?}
\begin{enumerate}[label=\Alph*.]
    \item Using the broadcast command instead of TTS
    \item Switching to Zigbee bulbs
    \item Turning on microphone sensitivity
    \item Downgrading firmware via USB
\end{enumerate}
\textit{Answer: A. One poster said they had to “resort to using broadcast commands which are clunky” when TTS stopped working.}

\item \textbf{What phrase did a Nest thermostat user shout after eco mode kept activating despite settings being disabled?}
\begin{enumerate}[label=\Alph*.]
    \item “Jeezus Google.”
    \item “Bravo Assistant!”
    \item “Mission accomplished!”
    \item “Danke Alexa.”
\end{enumerate}
\textit{Answer: A. The frustrated quote is “Jeezus Google,” showing irritation with unwanted eco behaviour.}

\item \textbf{When debating cloud versus local AI, which low-power device did a homeowner consider dedicating as an “always-on screen” for NotebookLM?}
\begin{enumerate}[label=\Alph*.]
    \item An old MacBook Pro
    \item A Raspberry Pi Zero
    \item A Lenovo Tab M8
    \item A Pixel 2 phone
\end{enumerate}
\textit{Answer: A. They planned “to dedicate an old MacBook Pro for the AI assistant” but were open to tablets.}

\item \textbf{Which free radio service did a listener compare to TuneIn, noting that Audacy retained signal ‘much better’ on Nest speakers?}
\begin{enumerate}[label=\Alph*.]
    \item Audacy
    \item Pandora
    \item SiriusXM
    \item iHeartWeather
\end{enumerate}
\textit{Answer: A. The poster said “The other services like Audacy work much better in terms of signal retention,” offering a comparison.}

\item \textbf{What is one reason a user said HomePods were a ‘let down’ compared with Google Nest speakers?}
\begin{enumerate}[label=\Alph*.]
    \item ‘You have to AirPlay from your phone’ for many tasks
    \item They lack any bass at all
    \item They cannot set alarms
    \item HomePods don’t support Wi-Fi
\end{enumerate}
\textit{Answer: A. The criticism: “on HomePod minis you have to air play from your phone… google are just way ahead,” indicating a comparison and feeling.}

\item \textbf{Faced with a 2025 Nest price hike e-mail, how many days of event video history does the ‘Nest Aware Plus’ plan promise to keep?}
\begin{enumerate}[label=\Alph*.]
    \item 60 days
    \item 30 days
    \item 7 days
    \item 120 days
\end{enumerate}
\textit{Answer: A. The notice states the plan keeps “60 days of event video history… up to 10 days of 24/7 video history,” summarising the aspect of subscription value.}

\end{enumerate}

\subsubsection{Google Nest Smart Speaker questionnaire from Qwen3-30B-A3B}
\label{subsection:Qwen3-30B-A3B questionnaire}
\begin{enumerate}

\item \textbf{In your smart home setup, how do you integrate Google Nest Mini speakers into an older home intercom system?}
\begin{enumerate}[label=\Alph*.]
    \item By replacing the entire intercom system with new wireless speakers
    \item By using a 24VDC power supply and buck converters to repurpose old wiring
    \item By connecting the speakers directly to a home Wi-Fi network without rewiring
    \item By mounting the speakers to the wall using adhesive pads only
\end{enumerate}
\textit{Answer: B. By using a 24VDC power supply and buck converters to repurpose old wiring}

\item \textbf{What is the primary reason for using Google Nest Audio speakers in a stereo setup for music playback?}
\begin{enumerate}[label=\Alph*.]
    \item They offer better sound quality and deeper bass than other smart speakers
    \item They are the only Google devices compatible with Apple Music
    \item They support Bluetooth audio output without delay
    \item They are designed specifically for use with Sonos systems
\end{enumerate}
\textit{Answer: A. They offer better sound quality and deeper bass than other smart speakers}

\item \textbf{When using Google Nest Mini as a temporary speaker during a move, what feature is most important for portability?}
\begin{enumerate}[label=\Alph*.]
    \item Built-in battery and rechargeable power source
    \item Ability to connect via USB-C to a TV
    \item Support for HDMI input from a computer
    \item Integration with Apple HomeKit for remote access
\end{enumerate}
\textit{Answer: A. Built-in battery and rechargeable power source}

\item \textbf{Why might a user choose Google Nest Mini over Apple HomePod Mini for a smart home setup?}
\begin{enumerate}[label=\Alph*.]
    \item Because Apple HomePod Mini requires a HomePod Hub for full functionality
    \item Because Google Nest Mini supports more third-party integrations and voice commands
    \item Because HomePod Mini lacks music playback capabilities
    \item Because Google Mini is compatible with all smart home ecosystems
\end{enumerate}
\textit{Answer: B. Because Google Nest Mini supports more third-party integrations and voice commands}

\item \textbf{What has caused frustration for users who rely on Google Nest speakers for music playback and voice commands?}
\begin{enumerate}[label=\Alph*.]
    \item Frequent unresponsiveness and inconsistent command recognition
    \item Inability to support Spotify Premium accounts
    \item Limited compatibility with Bluetooth devices
    \item High cost of replacing outdated models
\end{enumerate}
\textit{Answer: A. Frequent unresponsiveness and inconsistent command recognition}

\item \textbf{How do users perceive the reliability of Google Nest devices after the company’s shift to Gemini AI?}
\begin{enumerate}[label=\Alph*.]
    \item They believe the new AI will improve performance and voice recognition
    \item They are concerned about service degradation and potential obsolescence
    \item They expect all devices to be replaced within a year
    \item They trust that Google will maintain full support indefinitely
\end{enumerate}
\textit{Answer: B. They are concerned about service degradation and potential obsolescence}

\item \textbf{What do users suggest as the best way to maintain a smart home ecosystem without relying on Google’s services?}
\begin{enumerate}[label=\Alph*.]
    \item Using Home Assistant with a local server for full control
    \item Switching to Apple HomeKit for better integration
    \item Purchasing only Amazon Echo devices for compatibility
    \item Replacing all devices with non-smart alternatives
\end{enumerate}
\textit{Answer: A. Using Home Assistant with a local server for full control}

\item \textbf{What is a common workaround when a Google Nest speaker fails to connect to Wi-Fi during setup?}
\begin{enumerate}[label=\Alph*.]
    \item Changing the account region to the United States in the app settings
    \item Using a wired Ethernet connection instead of Wi-Fi
    \item Factory resetting the device and re-registering it
    \item Replacing the speaker with a newer model
\end{enumerate}
\textit{Answer: A. Changing the account region to the United States in the app settings}

\item \textbf{Why do some users prefer Google Nest Audio over other smart speakers for music listening?}
\begin{enumerate}[label=\Alph*.]
    \item It has superior stereo sound quality and bass response
    \item It is the most affordable smart speaker with no subscription fees
    \item It supports more music streaming services than any other model
    \item It integrates with Google Photos for visual album displays
\end{enumerate}
\textit{Answer: A. It has superior stereo sound quality and bass response}

\item \textbf{What is the main concern for users who are invested in both Apple and Google ecosystems?}
\begin{enumerate}[label=\Alph*.]
    \item Managing duplicate cloud storage between iCloud and Google One
    \item Choosing between Apple HomePod and Google Nest for smart home control
    \item Upgrading to new smartphones to maintain compatibility
    \item Finding compatible smart bulbs for both platforms
\end{enumerate}
\textit{Answer: A. Managing duplicate cloud storage between iCloud and Google One}

\item \textbf{How do users feel about the future of Google Nest hardware following recent service issues?}
\begin{enumerate}[label=\Alph*.]
    \item They believe Google will soon release a new generation of devices
    \item They are skeptical about long-term support and product continuity
    \item They expect Google to fully open-source the software for user maintenance
    \item They are confident that all devices will continue to function indefinitely
\end{enumerate}
\textit{Answer: B. They are skeptical about long-term support and product continuity}

\item \textbf{What challenge arises when trying to control Google Nest devices from a non-Google ecosystem device?}
\begin{enumerate}[label=\Alph*.]
    \item Voice commands are not recognized due to language mismatch
    \item The speaker cannot detect voice input from Apple devices
    \item Integration with Home Assistant becomes unreliable or impossible
    \item The device refuses to connect to Wi-Fi networks outside the Google network
\end{enumerate}
\textit{Answer: C. Integration with Home Assistant becomes unreliable or impossible}

\item \textbf{What is a key limitation when using Google Nest Mini speakers for doorbell alerts?}
\begin{enumerate}[label=\Alph*.]
    \item The alert only plays on one speaker, not all connected devices
    \item The doorbell chime only works when the speaker is unplugged
    \item The speaker cannot broadcast alerts during music playback
    \item Alerts are delayed by 10–15 seconds due to app routing
\end{enumerate}
\textit{Answer: A. The alert only plays on one speaker, not all connected devices}

\item \textbf{Why might a user consider replacing a Google Nest Hub Max despite its popularity?}
\begin{enumerate}[label=\Alph*.]
    \item It is no longer available for purchase in many regions
    \item It has a built-in camera that cannot be disabled
    \item It requires a paid subscription for basic functionality
    \item It only works with Apple devices and not Android
\end{enumerate}
\textit{Answer: A. It is no longer available for purchase in many regions}

\item \textbf{How do users describe the current performance of Google Nest Hub devices in 2025?}
\begin{enumerate}[label=\Alph*.]
    \item They are slow, glitchy, and frequently misinterpret commands
    \item They have improved dramatically with the new Gemini update
    \item They function as well as they did five years ago
    \item They are only useful for playing music and showing photos
\end{enumerate}
\textit{Answer: A. They are slow, glitchy, and frequently misinterpret commands}

\item \textbf{What is a major reason users are hesitant to expand their Google Nest ecosystem?}
\begin{enumerate}[label=\Alph*.]
    \item Fear of being locked into a system that may no longer be supported
    \item High cost of additional smart home accessories
    \item Incompatibility with older Wi-Fi routers
    \item Lack of support for non-English languages
\end{enumerate}
\textit{Answer: A. Fear of being locked into a system that may no longer be supported}

\item \textbf{What is the most common solution users suggest for improving audio performance across multiple Nest speakers?}
\begin{enumerate}[label=\Alph*.]
    \item Upgrading to a higher-end model like the Google Home Max
    \item Using a dedicated audio amplifier for better sound quality
    \item Replacing all speakers with Sonos devices for consistent playback
    \item Using a local server to manage audio streams and reduce latency
\end{enumerate}
\textit{Answer: D. Using a local server to manage audio streams and reduce latency}

\item \textbf{What is a major drawback of using Google Nest Audio as a TV speaker?}
\begin{enumerate}[label=\Alph*.]
    \item It has a significant audio delay when connected via Bluetooth
    \item It lacks a 3.5mm input port for direct audio connection
    \item It cannot play video content from streaming apps
    \item It only works with Google TV devices
\end{enumerate}
\textit{Answer: A. It has a significant audio delay when connected via Bluetooth}

\item \textbf{What do users suggest as an alternative to Google Nest speakers for a private, local smart home system?}
\begin{enumerate}[label=\Alph*.]
    \item Using a Raspberry Pi with a local LLM for voice control
    \item Replacing all devices with Amazon Echo Dot models
    \item Switching to Apple HomePods for better privacy
    \item Using only non-smart speakers with manual controls
\end{enumerate}
\textit{Answer: A. Using a Raspberry Pi with a local LLM for voice control}

\item \textbf{How do users perceive the value of older Google Nest devices like the Nest Mini and Nest Audio?}
\begin{enumerate}[label=\Alph*.]
    \item They are still functional and affordable, especially when bought secondhand
    \item They are outdated and no longer supported by Google
    \item They are only useful for basic tasks like playing alarms
    \item They are incompatible with modern Wi-Fi networks
\end{enumerate}
\textit{Answer: A. They are still functional and affordable, especially when bought secondhand}

\end{enumerate}

\subsection{Lighted Tree of the questionnaires on Google Nest Smart Speaker}
\label{subsection:lighted tree}
\begin{figure*}[h]
    \centering
    \includegraphics[width=1\linewidth]
    {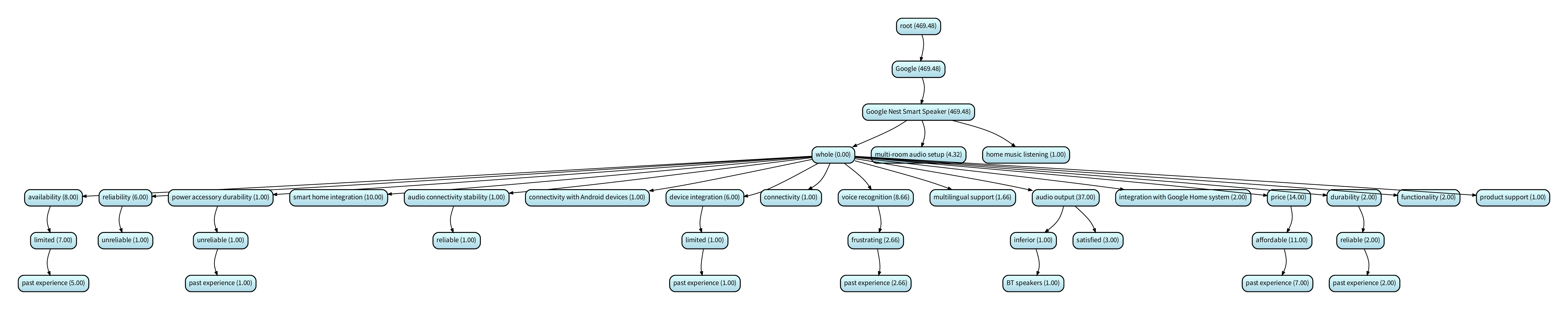}
\caption{Lighted Tree from GPT-5 on Google Nest Smart Speaker.}
    \label{fig:lighted-gpt-5}
\end{figure*}

\begin{figure*}[h]
    \centering
    \includegraphics[width=1\linewidth]
    {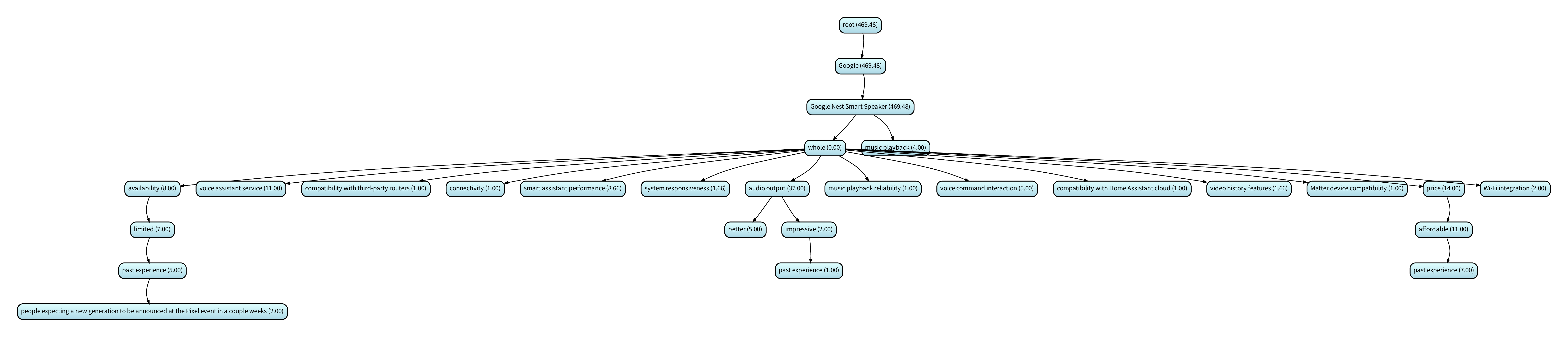}
\caption{Lighted Tree from GPT-o3 on Google Nest Smart Speaker.}
    \label{fig:lighted-gpt-o3}
\end{figure*}

\begin{figure*}[h]
    \centering
    \includegraphics[width=1\linewidth]
    {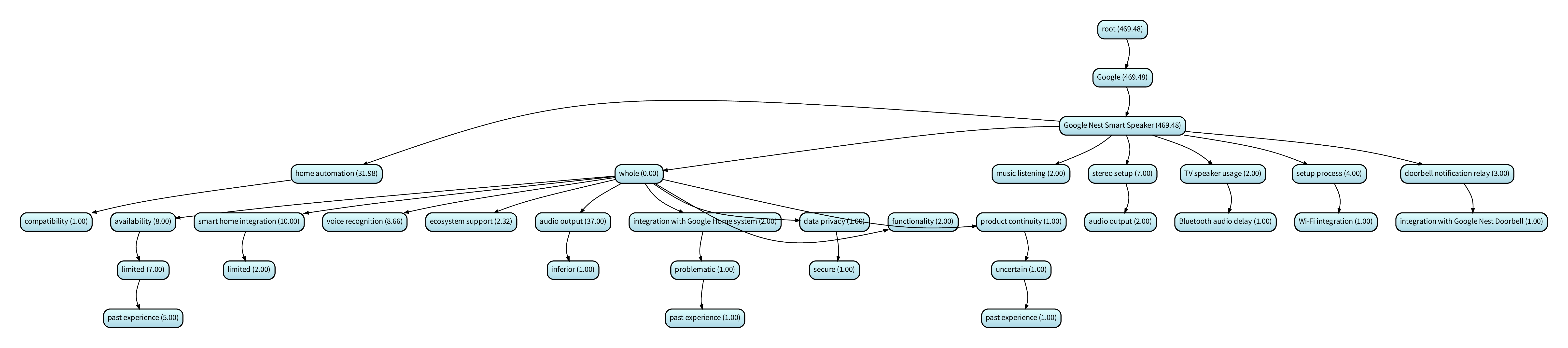}
\caption{Lighted Tree from Qwen3-30B-A3B on Google Nest Smart Speaker.}
    \label{fig:lighted-qwen3-30b-a3b}
\end{figure*}

%% file: Fig/data.tex
\begin{figure}
    \centering
    \includegraphics[width=0.5\textwidth]{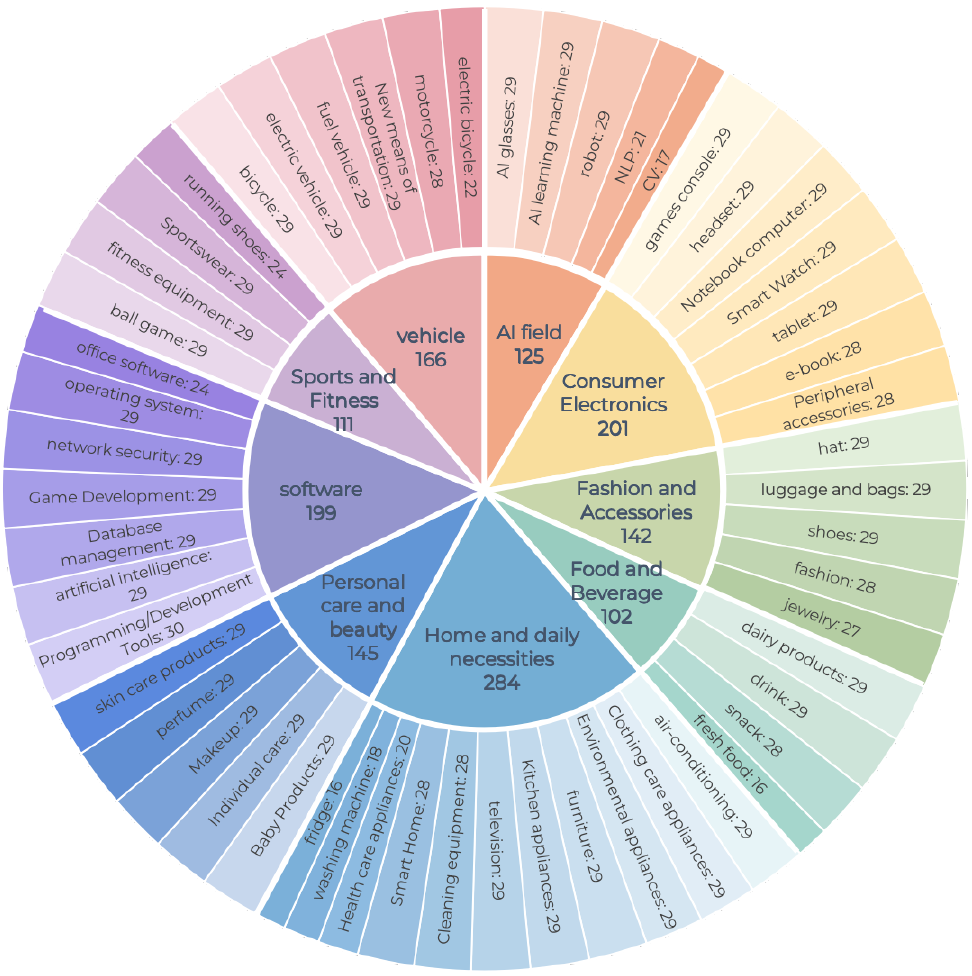}  
    \caption{Overview of \bench: It includes over 200k product-level discussions across 9 major domains, 54 sub-domains, and more than 1,400 products.}
    \label{fig:data}
\end{figure}